\documentclass[twocolumn,showpacs,preprintnumbers,prb,amsmath,amssymb]{revtex4}

\usepackage{graphicx}
\usepackage{dcolumn}
\usepackage{bm}

\begin{document}

\preprint{PRB-MgB2 Single crystal}

\title{Microstructure and pinning properties of hexagonal-disc shaped single
crystalline MgB$_{2}$}

\author{C. U. Jung}
 \altaffiliation[New address: ]{Tokura Spin Superstructure Project,
 ERATO, JST, AIST, Tsukuba Central 4
 Tsukuba, Ibaraki 305-8562, Japan}
 \email[E-mail: ]{cu-jung@aist.go.jp}
 \homepage[Url: ]{http://unit.aist.go.jp/cerc/}
\author{J. Y. Kim}
\author{P. Chowdhury}
\author{Kijoon H. P. Kim}
\author{Sung-Ik Lee}
\email[E-mail: ]{silee@postech.ac.kr}
 \homepage[Url: ]{http://www-psc.postech.ac.kr}
\affiliation{National Creative Research Initiative Center for
Superconductivity and Department of Physics, Pohang University of
Science and Technology, Pohang 790-784, Republic of Korea}
\author{D. S. Koh}
\affiliation{Department of Physics, Pohang University of Science
and Technology, Pohang 790-784, Republic of Korea}
\author{N. Tamura}
\author{W. A. Caldwell}
\affiliation{Lawrence Berkeley National Laboratory, Advanced Light
Source, 1 Cyclotron Road, MS-2-400 Berkeley, CA 94720, USA}
\author{J. R. Patel}
\affiliation{Lawrence Berkeley National Laboratory, Advanced Light
Source, 1 Cyclotron Road, MS 7-222 Berkeley, CA 94720, USA, and
SSRL/SLAC, Stanford University, CA 94309, USA}

\date{\today}

\begin{abstract}
We synthesized hexagonal-disc-shaped MgB$_{2}$ single crystals
under high-pressure conditions and analyzed the microstructure and
pinning properties. The lattice constants and the Laue pattern of
the crystals from X-ray micro-diffraction showed the crystal
symmetry of MgB$_{2}$. A thorough crystallographic mapping within
a single crystal showed that the edge and c-axis of hexagonal-disc
shape exactly matched the (10-10) and the (0001) directions of the
MgB$_{2}$ phase. Thus, these well-shaped single crystals may be
the best candidates for studying the direction dependences of the
physical properties. The magnetization curve and the magnetic
hysteresis for these single crystals showed the existence of a
wide reversible region and weak pinning properties, which
supported our single crystals being very clean.
\end{abstract}

\maketitle

The recent discovery\cite{find} of superconductivity in MgB$_{2}$
has attracted great scientific\cite{KangHall,BudkoPRB} and
industrial \cite {Kangfilm,Larbal} interest. Even though basic
issues such as the carrier type \cite{KangHall} were addressed
immediately, conflicting reports still exist, especially on the
transport properties. For example, the residual resistivity ratio
(RRR) of bulk MgB$_{2}$ ranges from 1.2 to 30. \cite {JungNKim}

Higher quality bulk MgB$_{2}$ (called {\it highRRR-MgB}$_{2}$) has
been claimed to have a higher values of RRR $(20\sim 25)$, a low
residual resistivity ($\rho (40$ K$)<1$ $\mu \Omega $cm), a higher
magneto-resistance (MR), and a resistivity upturn at low
temperature under high magnetic field. \cite{BudkoPRB} Insulating
impurities and/or local strains have been suggested as possible
origins for these different observations.\cite
{Zhu,Chu1,Chu2,Chu3,0598} However, very recently, the existence of
unreacted Mg successfully explained the unusual enhancement of the
RRR and the MR in polycrystalline MgB$_{2}$.\cite{JungNKim}

Compared to the wide distribution of the RRR for polycrystals, the
RRR in the ab-plane resistivity for single crystals has a narrow
distribution.\cite {Lee,Xu} The RRR of single crystals ranges from
5 to 6, which is much smaller than the values for {\it
highRRR-MgB}$_{2}$ bulk samples, but similar to the values for
well-prepared polycrystals.\cite
{JungNKim,Nesterenko1,Nesterenko2,Fuchs} The anisotropy factor
$(\gamma \sim 3)$ and the Debye temperature $(\Theta \sim 1160\pm
60$ K) are also consistent among various reports.\cite{MR}

It is interesting to note that the magnetic properties are
somewhat different for different single
crystals.\cite{Lee,Xu,Lima} The superconducting transition width
in the zero-field-cooled magnetization for single crystals as
large as a few hundred $\mu $m is a little bit broader than
expected. The ratio of the low-field magnetization in the
field-cooled (FC) state to that in the zero-field-cooled (ZFC)
state gives a rough indication of the pinning strength, and this
ratio is quite different for different single crystals. The
magnetic hysteresis for single crystals having very weak pinning
can be used to probe impurities. In one report, the magnetic
hysteresis contained\ a significant amount of paramagnetic
component. In another report, the magnetic hysteresis for an
aggregation of single crystals showed a large ferromagnetic
contribution and a significant irreversible
magnetization.\cite{Lima}

For the above reasons, it is necessary to confirm the quality of
single crystals in more detail before performing the main
measurements. The existence of impurities and structural
imperfections on a microscopic scale can result in diverse
physical properties. Here, we report the growth of, as well as
X-ray micro-diffraction and magnetization measurements for,
MgB$_{2}$ single crystals with hexagonal-disc shapes and shiny
surfaces. Our single crystals are unique as far as the shape is
concerned. The diagonal length
and the thickness for the largest crystal was about 100 $\mu $m and 10 $\mu $%
m, respectively. The crystallinity was thoroughly identified by
using the Laue pattern in the X-ray micro-diffraction measurement.
Both the edge and the c-axis of the hexagonally shape disc were
found to match the crystal symmetry. The magnetization study
showed that pinning was very weak for our hexagonal-disc-shaped
single crystals.

Two different procedures were used to grow the single crystals,
and in both cases, excess Mg was critical for the growth of single
crystals. The first involved a two-step method in which already
synthesized pieces of MgB$_{2}$ bulk\cite{JungHPsynthesis} were
used as a seed material. They were heat treated in a Mg flux
inside a Nb tube, which was sealed in an inert gas atmosphere.
Then, the Nb tube was put inside a quartz tube, which was sealed
in vacuum. The quartz tube was heated for one hour at 1050$^{\circ
}$C, cooled very slowly to 700$^{\circ }$C for five to fifteen
days, and then quenched to room temperature. The temperature
dependences of the ab-plane resistivities for $H//ab$ and $H//c$
and high-resolution transmission electron microscope images have
already been reported for the single crystals synthesized using
this procedure.\cite{SingleKim}

\begin{figure}[tbp]
\includegraphics[width=7.0cm]{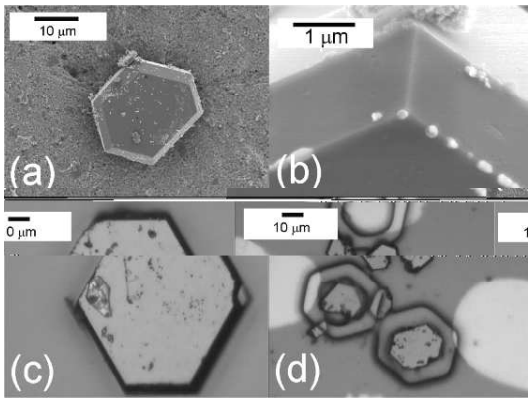}
\caption{\label{fig1} SEM and polarizing optical microscope images
of the MgB$_{2}$ single crystals with a hexagonal-disc shape. (a)
MgB$_{2}$ made in a two-step process with a diagonal distance of
about 25 $\protect\mu $m and with a thickness of about 4
$\protect\mu $m. (b) High-resolution image at the corner of
another crystal, and (c) polarizing optical microscope image of a
single crystal grown in a one-step process. (d) Polarizing optical
microscope images of MgB$_{2}$ single crystals. An epoxy was used
to fix six single crystals at the center of 100-$\protect\mu
$m-wide Cu crosshairs.}
\end{figure}

In the one-step method, 1 : 1 mixtures of Mg and B powders were
well ground and pressed into a pellet. Then, the pellet was placed
in a Ta capsule. This capsule was put in a high-pressure cell
equipped with a graphite heater. Heat treatment was done inside a
14-mm cubic multi-anvil-type press under 3
GPa.\cite{JungHPsynthesis} The heating temperature was around $1300^{\circ }$%
C. The temperature was maintained for about 30 minutes, and then
was slowly lowered to $800\sim 900^{\circ }$C. The final product
was a pellet containing a mixture of single-crystalline MgB$_{2}$
and Mg flux.

It was found that the one-step method usually gave larger
crystals, which was quite advantageous for the magnetization study
in this research. The crystal images were observed using a
polarizing optical microscope and a field-emission scanning
electron microscope (SEM). We successfully separated
single-crystalline MgB$_{2}$\ from the Mg flux by using a
thermo-mechanical spinning method. This method is possible due to
the fact that the melting (and/or decomposition) temperature of
MgB$_{2}$\ is higher than that of Mg.

Single crystals with sizes of tens of $\mu $m were selectively
handled by using a homemade micro-tweezers and were fixed on Si
substrates by using a photoresist as an epoxy. Since the volume of
one crystal made in the one-step process was still rather small,
we gathered about 200 single crystals on one substrate with their
{\it c}-axes aligned perpendicular to the substrate
surface.\cite{Lima} For the X-ray micro-diffraction measurements,
several crystals were fixed at the center of Cu crosshairs on the
substrate, as shown in Fig.~\ref{fig1}(d).\cite{SingleKim} The Cu
crosshairs was used for the only purpose to facilitate the
location of the sample by looking at the Cu fluorescence. The
magnetic properties were measured with a SQUID magnetometer
(Quantum Design, MPMS-{\it XL}).

The instrument used at the Advanced Light Source (ALS) for X-ray
micro-diffraction is capable of producing a submicron-size X-ray
microbeam and with submicron spatial resolution can probe the
local texture in a single crystal.\cite{NIMA} The sample was
positioned using the Cu fluorescence signal detected from the Cu
crosshairs on the Si substrate by using a high-purity Ge ORTEC
solid-state detector connected to a multichannel analyzer. The
crystal orientation with respect to the substrate can be
determined with an accuracy of 0.01 degree.

Figure\ \ref{fig1}(a) shows a typical SEM image for a MgB$_{2}$
single crystal synthesized in a two-step process. The crystals
observed by using a polarizing optical microscope had
hexagonal-disc shapes with edge angles of 120 degrees and very
flat and shiny surfaces. The sizes of the crystals were about
$20\sim 60$ $\mu $m in diagonal length and $2\sim 6$ $\mu $m in
thickness. Figure\ \ref{fig1}(b) shows a magnified view of the
upper corner of another crystal. The smooth surfaces and the sharp
edges confirm that our small crystals had a very low probability
of having mosaic aggregates of nanocrystals either along the
$ab$-plane or along the $c$-axis; thus, we had a better chance to
study their intrinsic properties. The single crystals reported so
far, except for those reported by us, have had irregular shapes.
\cite{Lee,Xu,Lima}

Figure \ref{fig1}(c) shows a polarizing optical microscope image of a MgB$%
_{2}$ single crystal synthesized in a one-step process, which
resulted in larger crystals. The sizes of these crystals were
$30\sim 100$ $\mu $m in diagonal length, so they were picked for
measurements of the magnetic properties and the X-ray
micro-diffraction.

\begin{figure}[tbp]
\includegraphics[width=8cm]{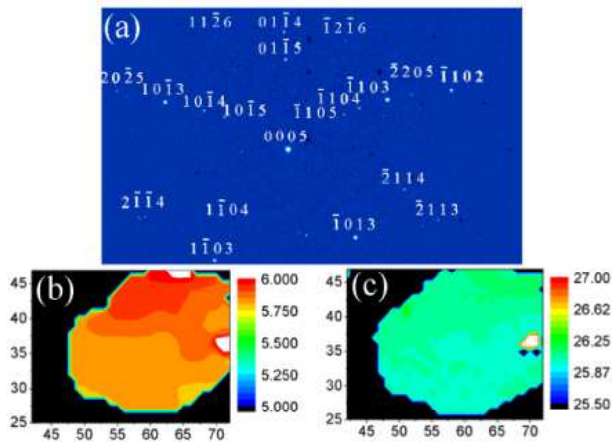}
\caption{\label{fig2} (a) A representative image of an indexed
Laue pattern from X-ray micro-diffraction. (b) and (c) are,
respectively, for the out-of-plane and the in-plane orientations
inside a single crystal. The out-of-plane orientation showed a
variation of about 0.2 degrees between the bottom and the top
parts, indicating a slight bending of the crystal. The in-plane
orientation also showed slight inhomogeneities of up to about 0.2
degrees.}
\end{figure}

The crystal structure was identified by using white beam X-ray
micro-diffraction measurements. After positioning these single
crystals on the substrate in Fig.~\ref{fig1}(d), a $100$ $\mu $m
$\times 100$ $\mu $m region between the Cu crosshairs was scanned
with a step size of 2 $\mu $m. At each step, the Laue pattern
(together with the Cu {\it K} fluorescence signal ) was collected
with a BRUKER 6000 CCD camera which has an active area of $9\times
9$ cm and was placed about 4 cm above the sample. [2500 images,
1024 pixel $\times 1024$ pixel mode] The exposure time at each
step, was 1 second. An example of a Laue pattern obtained from a
MgB$_{2}$ single crystal is shown in Fig.~\ref{fig2}. The Laue
patterns are consistent with a hexagonal MgB$_{2}$ structure
($a=3.086$ \AA , $c=3.524$ \AA , Space Group number =191, Ref. 1).
The silicon reflections from the substrate were digitally removed.
Typically, more than 20 reflections with energies ranging from 5
to 14 keV were indexed successfully. The (0005) reflection in the
center of the pattern in Fig.~\ref{fig2} corresponds to the
direction of the normal to the crystal surface. This confirms that
the surface plane normal is along the c-axis. Moreover, the
hexagonal edges of the crystals were found to match the
\mbox{$<$}%
1,0,-1,0%
\mbox{$>$}%
directions within a fraction of a degree resolution. Thus, the
shapes of the crystals in the microscope image followed the
MgB$_{2}$\ crystal symmetry, which will be quite useful for any
research of the direction dependencies of the physical properties
in MgB$_{2}$.

Indexing the Laue patterns in Fig.~\ref{fig2}(a) allowed us to
calculate the complete orientation matrix of the X-ray illuminated
volume. A finer step size of 1 $\mu $m was used for the white-beam
scan. The orientation variations inside the single crystal shown
in the right bottom corner of Fig \ref{fig1}(d) are shown in Figs.
\ref{fig2}(b) and \ref{fig2}(c). Figure \ref {fig2}(b) is the
out-of-plane orientation variation calculated as the angle between
the c-axis and the normal to the surface of the silicon substrate.
The out-of-plane variation was about 0.2 degrees between the light
orange and the red regions. The out-of-plane orientation shows a
variation of about 0.2 degrees between the bottom and the top
parts, indicating a slight bending of the crystal (which might be
due to the photoresist used as an epoxy). Figure \ref{fig2}(c)
shows the in-plane orientation variation calculated as the angle
between the measured a-axis (or b-axis) and a reference
directions. The variation was about 0.4 degrees between the light
blue-green and the green regions. The in-plane orientation also
showed some inhomogeneities of up to about 0.2 degrees.

\begin{figure}[tbp]
\includegraphics[width=5.0cm]{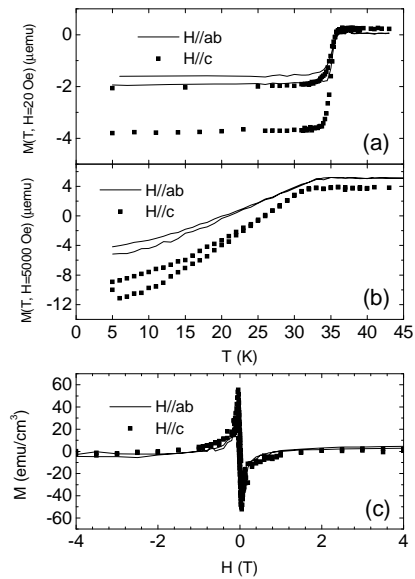}
\caption{\label{fig3} (a) Low-field magnetization curve $M$($T$) measured at 20 Oe for $%
H//c$ and $H//ab$. (b) Magnetization curves measured at 5000 Oe
showed a wide reversible region. $T_{{\rm irr}}(H=5000$ Oe) was 26
and 24 K for $H//c$ and $H//ab$, respectively. (c) The magnetic
hysteresis $M$($H$) at 5 K; the symbols and the line are for
$H//c$ and $H//ab$, respectively.}
\end{figure}

These results demonstrate that the orientation of crystal axis of
our hexagonal-disc-shaped single crystals was perfect, within 0.2
degrees. A recent study showed that (0001) twist grain-boundaries,
formed by rotations along the $c$-axis (typically by about 4
degrees), were the major grain boundaries in polycrystalline
MgB$_{2}$.\cite{Zhu} This kind of grain boundary was attributed to
the weaker Mg-B bonding.\cite{Zhu}

To study the bulk nature of the superconductivity, we measured the
magnetization curve $M$($T$) and the magnetic hysteresis $M$($H$).
Figure \ref{fig3}(a) shows the magnetization curve $M$($T$)\
measured at 20 Oe in the ZFC and the FC states. The $T_{c}$ onset
was about $38$ K. The FC signals for fields parallel and
perpendicular to the c-axis were larger than 60\% of the ZFC
signals, which suggested that the pinning was very weak. The
overall shape was not much different for these two orientations.
The different values of the magnetic moment for the different
field directions were due to a demagnetization effect caused by
the planar-disc shapes of the crystals.\cite{SingleKim} The
transition width in the $M(T)$ curve was much narrower than that
for single crystals made in the two-step method.\cite {SingleKim}
The volumes of the crystals made in the one-step process were
about one order of magnitude larger than the volumes of the
crystals made in the two-step process.

The weak pinning was demonstrated by the magnetization at 5000 Oe
and the magnetic hysteresis $M$($H$) at 5 K, as shown in
Fig.~\ref{fig3}(b) and (c), respectively. With
$(M(H^{-})-M(H^{+}))=0.1$ emu/cm$^{3}$ as the criterion for the
reversible point, the values of $T_{{\rm irr}}(H=5000$ Oe) were 26
and 24 K for $H//c$ and $H//ab$, respectively. At each
temperature, the upper critical fields obtained from the
resistivity measurements were about 1.7 and 6 Tesla for our
well-shaped single crystals.\cite{SingleKim} Thus, a
very wide reversible region existed in our single crystals, especially for $%
H//ab$. [$H_{{\rm irr}}$ for bulk MgB$_{2}$ at $T=25$ K was about
3 T\cite
{Fuchs}, which is about 6 times higher than that of single crystalline MgB$%
_{2}$. ]

The weak pinning shown by the wide reversible region was also
consistent with the results of the $M(H)$ measurement. The $M(H)$
at 5 K in Fig.~\ref{fig3}(c) showed negligible paramagnetic or
ferromagnetic background up to 5 Tesla. Reversible magnetization
was dominant for the $M(H)$ of our single crystals, which was
quite consistent with fact that the pinning in our single crystals
was weak and with the existence of wide reversible region
demonstrated by the lower $H_{{\rm irr}}$ value in Fig
\ref{fig3}(b). The different slopes of the $M(H)$ curves at the
starting low fields ($H<300$ Oe) for the different field
directions were due to different demagnetization factors.

Previously, the $M(H)$ for irregular shaped single crystals was
shown to have a non-negligible irreversible
contribution\cite{Lima} and a quite large paramagnetic and/or
ferromagnetic background signal, which might have been due to the
existence of impurities.\cite{Xu,Lima} The results in Fig.\ \ref
{fig3} indicate that the strong bulk pinning previously reported
for polycrystalline \cite{kim} and thin films\cite{Kangfilm} might
be due to entirely extrinsic pinning sites, such as grain
boundaries and crystallographic defects.\cite{Pissas} This is
consistent with the absence of core pinning, even at $T\sim
0.5\times T_{c}$, for bulk sample.\cite {Larbal}

In summary, we report the structural and magnetic properties of
MgB$_{2}$ single crystals with hexagonal-disc shapes. The X-ray
micro-diffraction showed that the hexagonal-disc shape of the
single crystal followed the crystallographic symmetry, which will
be very useful for studying orientation-dependent physical
properties. The magnetization curve and the magnetic hysteresis
provided consistent evidence that our single crystals were very
clean and had very weak pinning: the large FC magnetization and
the sharp transition of the ZFC magnetization at low field $M(T)$,
the wide reversible region, and the $M$($H$, $T=5$ K) dominated by
the reversible magnetization without any significant paramagnetic
and/or ferromagnetic contribution.

\begin{acknowledgments}
We appreciate M. H. Kim and S. R. Jung for their help in handling
of tiny crystals. This work is supported by the Ministry of
Science and Technology of Korea through the Creative Research
Initiative Program. The Advanced Light Source is supported by the
Director, Office of Science, Office of Basic Energy Sciences,
Materials Sciences Division, of the U.S. Department of Energy
under Contract No. DE-AC03-76SF00098 at Lawrence Berkeley National
Laboratory.
\end{acknowledgments}


\begin{references}
\bibitem{MR}  However, even in single crystals, some spectra for the
transport properties still exist. MR values at 5 T ranged from
about 2\% to 20\% and the MR ratio for the two magnetic field
directions shows a wide variation among reports.
\end{references}
\end{document}